# Measurements of Radon Concentrations Using CR-39 Detectors in China JinPing Underground Laboratory (2015-2016)

Cuihong Liu[a], Hao Ma [a], Zhi Zeng [a,*], Jianping Cheng[a,b], Junli Li[a], Hui Zhang[a]

a. Key Laboratory of Particle and Radiation Imaging (Ministry of Education) and Department of Engineering Physics, Tsinghua University, Beijing 100084, China;
b. College of Nuclear Science and Technology, Beijing Normal University, Beijing 100875

*corresponding author: zengzhi@tsinghua.edu.cn（Zhi Zeng）

## Abstract:

Radon background is one of the most critical influences on the ultra-low background and rare-event physical experiments in underground laboratories. To conduct a comprehensive investigation of the radon concentrations in China JinPing underground Laboratory (CJPL), long-term cumulative measurements were carried out using time-integrated passive radon dosimeters containing CR-39 solid-state nuclear track detectors at four sites of CJPL-I. The measurements lasted from May 30,2015 to March 16,2016, a total of 290 days, and the total effective measurement time is 6,953 hours. In the main experiment hall equipped with the ventilation system, the average annual radon concentrations were (55±7) Bq·m$^{-3}$, (58±10) Bq·m$^{-3}$ and (53±9) Bq·m$^{-3}$ in three locations respectively. In the storage tunnel without any radon reduction facility, the average annual radon concentrations increased to (345±15) Bq·m$^{-3}$, reflecting the original average annual radon concentration levels in the underground laboratory. Compared with the measured results in the period from 2010 to 2011, the average radon concentration in the main experimental hall was reduced from (86±25) Bq·m$^{-3}$ to (58±10) Bq·m$^{-3}$ due to the well-running ventilation system. Compared with other international underground laboratories, radon concentrations in CJPL-I were at an intermediate level and can meet the demand for low background and rare-event physical experiments.

## 1. Introduction

China JinPing underground Laboratory (CJPL) is the deepest underground laboratory in the world located in the middle of the traffic tunnel through Jinping Mountain in southwest China(Cheng et al., 2017; Hu et al., 2017; Mi et al., 2015). With 2400-meter rock overburden, the cosmic ray flux in CJPL is so low that it is a perfect underground laboratory for ultra-low background and rare-event physical

experiments(Wu et al., 2013). For these experiments, their sensitivity largely depends on the recognition and reduction of the background induced by cosmic rays, gammas, neutrons and radioactive gas radon originating from the surrounding environment. Among them, the contribution of radioactive gas radon and its progenies is significant since it is widely present in rocks which are the main construction materials of the underground laboratory(Udovičić et al., 2009).To reduce radon concentration, many radon reduction methods were implemented in CJPL, such as establishing the effective ventilation system to bring in fresh air(Cheng et al., 2017; Mi et al., 2015), brushing the anti-radon coating to minimize radon exhalation, cleaning the partial sealed shield with nitrogen, etc. To assess the effectiveness of these radon reduction methods and evaluate the effect of radon background on experiments, continuous monitoring of radon concentrations has been carried out in CJPL-I using two AlphaGUARDPQ2000 radon monitors since the establishment of the laboratory in December 2010.The measurement results were summarized in Ref (Cheng et al., 2017).So far, the laboratory has been running for seven years and the ventilation system is working well. To conduct an overall investigation of radon concentrations in CJPL, more monitoring methods have been applied since May 30, 2015, including cumulative measurement and instantaneous measurement. In this paper, we will detail the cumulative measurements from May 30,2015 to March 16,2016.

## 2. Materials and Methods

### 2.1. CR-39 Detection Principle

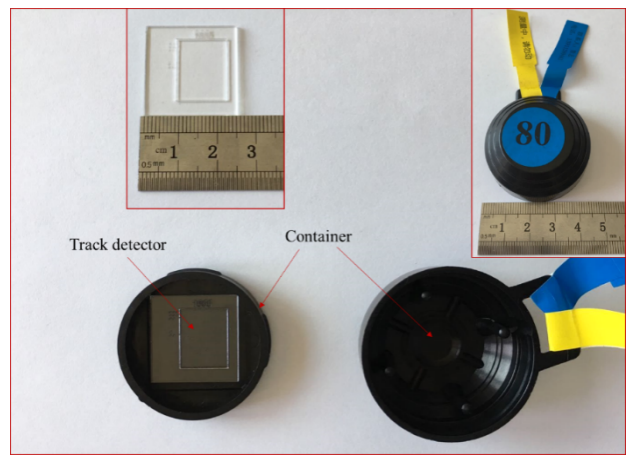

Fig.1.CR-39 nuclear track detector and its container used in CJPL-I

To measure radon cumulative concentration, time-integrated passive radon dosimeters containing CR-39 solid-state nuclear track detectors were used in this study, as shown in Fig.1. They were purchased from Track Analysis System Ltd (TASL, UK). The selected CR-39 detector is made of polyallyl diglycol carbonate with a thickness of 1.5mm. Its effective area is 2.38cm$^2$ and can be fixed in the dosimeter's groove. There are some slits around the dosimeter that allow radon gas to diffuse into its chamber and prevent other contaminants. The alpha particles generated from the decay of radon and its daughters in the air volume of the chamber can interact with the CR-39 film, causing a damage to the molecular bonds of the polymer. The damage manifests itself as sub-microscopic damage tracks on the surface of the film. Under certain chemical or electrochemical etching conditions, these latent tracks can be enlarged to observable permanent tracks. By using an optical amplification readout device, the number of these tracks can be calculated and the equilibrium radon concentration can be given by the following equation:

$$C_{Rn} = \frac{N_T - N_B}{AkT} \tag{1}$$

Where $C_{Rn}$ is the equilibrium radon concentration in Bq·m$^{-3}$, $N_T$ is the registered whole tracks in T$_r$ in the CR-39 detector, $N_T$ is the background tracks in detectors used for measurement in T$_r$, A is the effective area of the CR-39 detector in cm$^2$, $k$ is the calibration factor of CR-39 detector in T$_r$·cm$^{-2}$ /kBq·m$^{-3}$·h and T is the effective exposure time in hours. The calibration factor $k$ (T$_r$·cm$^{-2}$ /kBq·m$^{-3}$·h) is used to convert the net track density to the value of radon measurement and should be determined in the standard radon chamber before radon measurements. It can be calculated using the following equation:

$$k = \frac{N_0 - N_{B0}}{AC_0} \tag{2}$$

Where $C_0$ is the integrated concentration in kBq·m$^{-3}$·h in the standard radon chamber, $N_0$ is the irradiated tracks in Tr, $N_{B0}$ is the background tracks in detectors used for calibration in Tr and T is the effective exposure time in hours.

From equation (1) and (2), the equilibrium radon concentration can be rewritten by the following summarized equation:

$$C_{Rn} = \frac{(N_T - N_B) C_0}{(N_0 - N_{B0}) T} = \frac{N_{net} C_0}{N_{net0} T} \tag{3}$$

Where $N_{net}$ is the net track during the measurement, and $N_{net0}$ is the net track during the calibration? As can be observed in equation (3), uncertainties in radon concentration measurements were from: background tracks, irradiated tracks, exposure time and detector calibration factors, including the integrated concentration in the standard radon chamber and corresponding their radiated tracks and the background tracks. Therefore, the combined standard uncertainty of measured radon concentrations can be calculated with the following function:

$$U_C^2(c_{Rn}) = \left(\frac{\partial C_{Rn}}{\partial N_{net}}\right)^2 u(N_{net})^2 + \left(\frac{\partial C_{Rn}}{\partial C_0}\right)^2 u(C_0)^2 + \left(\frac{\partial C_{Rn}}{\partial N_{net0}}\right)^2 u(N_{net0})^2 + \left(\frac{\partial C_{Rn}}{\partial T}\right)^2 u(T)^2$$

$$= \left(\frac{C_{Rn}}{N_{net}}\right)^2 u(N_{net})^2 + \left(\frac{C_{Rn}}{C_0}\right)^2 u(C_0)^2 + \left(\frac{C_{Rn}}{N_{net0}}\right)^2 u(N_{net0})^2 \tag{4}$$

Where $U_C(c_{Rn})$ is the combined standard uncertainty of measured radon concentrations in Bq·m$^{-3}$, $u(N_{net})$ is the uncertainty of net tracks in the CR-39 detector used for measurements in T$_r$, $u(N_{net0})$ is the uncertainty of net tracks in the CR-39 detector used for calibration in Tr and $u(C_0)$ is the uncertainty of the integrated concentration in kBq·m$^{-3}$·h in the standard radon chamber. The uncertainty of the exposure time $u(T)$ is relatively low and negligible (Friedmann et al., 2017; Pantelić et al., 2014). The $u(N_{net})$ and $u(N_{net0})$ can be calculated with the following functions:

$$u(N_{net})^2 = u(N_T)^2 + u(N_B)^2 \tag{5}$$

$$u(N_{net0})^2 = u(N_0)^2 + u(N_{B0})^2 \tag{6}$$

Among them, uncertainties of the background tracks were derived from the inherent defects of the CR-39 film, cosmic ray bombardment and poor sealing during transportation and storage. The uncertainties of the irradiated tracks mainly came from the readout system and the etching process (Al-Khateeb et al., 2017)which included the configuration of etching solution, control of etching temperatures and etching duration. The uncertainties for radon concentrations in the standard chamber were mainly due to the uncertainties of radon sources, radon decay constant, the calibration factor of the continuous monitor, etc.(Abo-Elmagd et al., 2018; Chakraverty et al., 2018).

From the above discussion, it can be seen that the accuracy of the etched tracks, which is closely related to etching conditions, and the calibration factor are important for the quality of radon concentration measurement. Therefore, in this study prior to radon measurements, the etching conditions were optimized and the detector calibration factor was determined using the standard radon chamber.

## 2.2. Etching Conditions Optimization

Before the chemical etching of irradiated detectors, three parameters need to be determined, including the etching solution concentration, the etching temperature and the duration of the etching. Due to the inherent characteristics of the CR-39 detector, there is no general set of etching conditions for all films. There are differences in detectors produced by different manufactures, even different batches from the same manufacturer. To obtain the optimal etching conditions for the detectors used in this study, a series of experiments were conducted. The etching solution was fixed in a 6.25 mol/L sodium hydroxide (NaOH) aqueous solution. Different combinations of etching temperature and etching duration were selected without changing the etching solution concentration. All etched detectors were irradiated in the standard radon chamber at the same radon concentration level. According to the etching efficiency and characteristics of the etched tracks, including density, shape (circular, elliptical, wedge), clarity (easy to identify) and area density, the etch results were compared and the optimized etching parameters of this batch of detectors were determined.

## 2.3. Detector Calibration

The detector calibration factor is a key parameter for conversion between radon concentrations and net tracks densities. To obtain more accurate factors, a series of calibration experiments were conducted in standard radon chamber at National Institute of Metrology in China. A batch of diffusion chambers with CR-39 detectors were simultaneously placed in the radon chamber and three levels of accumulated radon concentrations were selected. Throughout the experiment, radon concentrations in the chamber were continuously monitored by the calibrated commercial AlphaGUARDPQ2000（Genitron Instruments GmbH, Germany）radon detector based on an optimized-design ionization chamber and the environmental parameters remained unchanged. After exposure, the detectors were collected and etched together under the optimum etching conditions as described above. The etched tracks were counted by the readout system and the calibration factor was calculated by equation (2).

## 2.4. The Distribution of CR-39 Detectors in CJPL-I

CJPL-I has a total volume of 4000 m$^3$(Cheng et al., 2017). As shown in Fig.2, the laboratory is divided into three parts: the entrance tunnel, the connection tunnel and the main experimental hall. The main experimental hall has dimensions of 6.5m (width)×6.5m(height) ×40m(length) and consists of the polyethylene shielding (PE) room and the main hall with two floors(Cheng et al., 2017).The bedrock surrounding CJPL is made of marble and the laboratory space is shielded from the bare rock by a 0.5-m-thick layer of concrete. The radon exposure measurements were carried out in the main experimental hall and the storage tunnel located in the corner of the preparation tunnel (see Fig.1). In the experimental hall, the ventilation system was equipped with a 10kmlong pipe with a diameter of 55 cm (Cheng et al., 2017). Fresh air outside the tunnel can be pumped into the laboratory at an air-exchange rate of 4500m$^3$h$^{-1}$. The storage tunnel was mainly used to store sundries without any ventilation equipment.

For long-term cumulative radon measurements, approximately 48 diffusion chambers with CR-39 nuclear track detectors were distributed at four sites, including the PE room, the main hall, the second floor of the hall and the storage tunnel. The schematic layout of the measurement is shown in Fig.2. At each site, 12 samples were hung or placed at a distance of 1.5 to 2 meters above the floor and two pairs of parallel samples were deployed. To further investigate the spatial variations of radon concentrations, the samples were distributed almost uniformly at each site. In addition, 5 detectors are left as blank samples corresponding to the total number of samples.

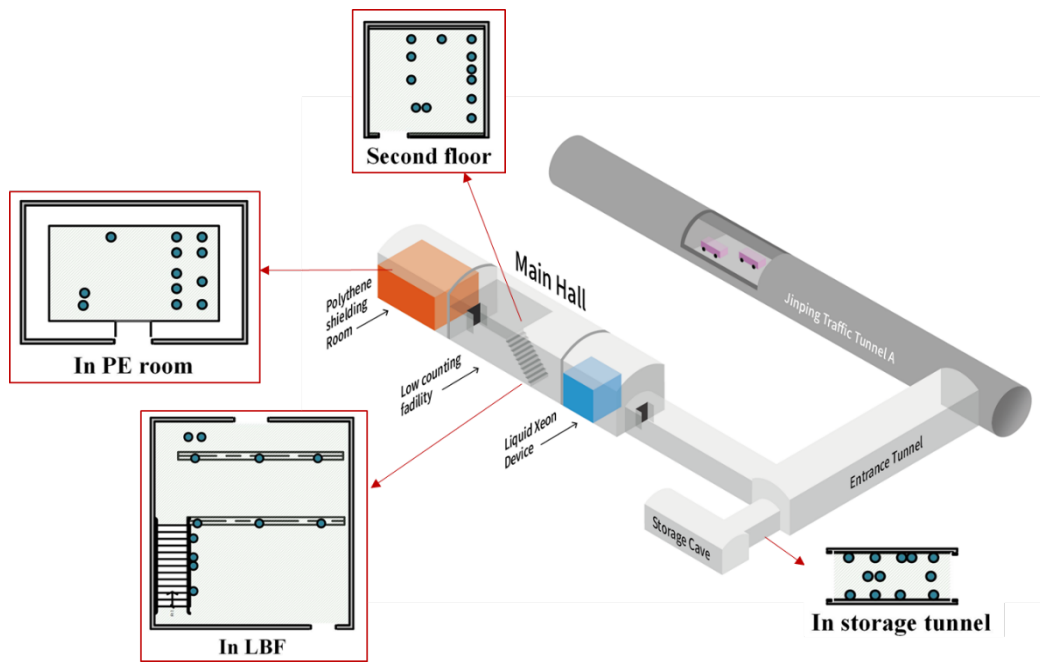

(a)

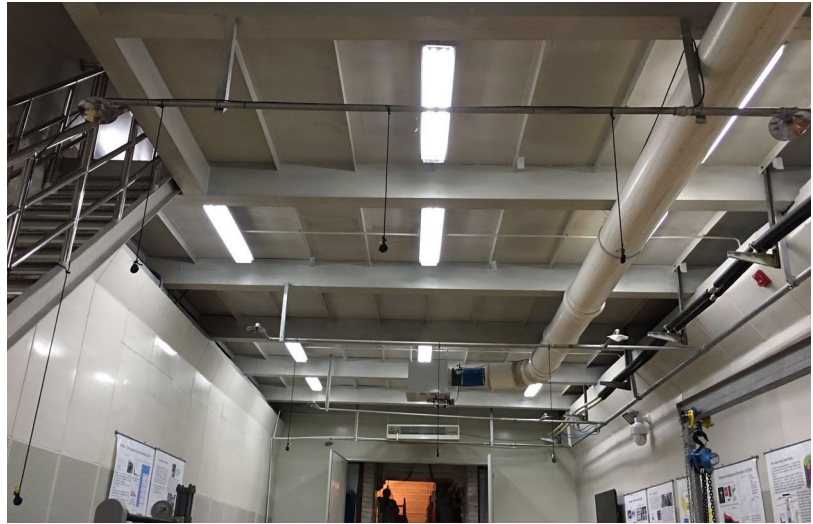

(b)

Fig.2. The layout scheme of the measurement points in CJPL-I. (a)The layout of CJPL-I(b) A realistic image of detector distribution in the main hall in CJPL-I. The measurements were conducted in four green parts and dots represented the sampling points.

## 2.5. Measurements Distribution in CJPL-I

The measurements lasted from May 30$^{th}$2015 to March 16$^{th}$2016, a total of 290 days, and the total effective measurement time is 6,953 hours. After exposure, the detectors were collected and sealed to isolate the external environment and transported to the laboratory in Tsinghua University for processing and analysis. In the laboratory, all detectors were carefully removed from the chambers and chemically etched under the optimum etching conditions as described above. To minimize accidental errors due to little variation in temperature and concentration of the etching solution, all irradiated detectors as well as blank background detectors were etched simultaneously. After etching, a series of neutralization, cleaning and drying operations were carried out to clean the detectors. The tracks in the etched detectors were automatically scanned, identified and counted using the *TASLIMAGE* evaluation system, which consists of a motorized scanning platform with three-axis controlled by a software, an apochromatic optical microscope with high-quality Nikon optics, a scanning (etching) frame, an analysis control software and a dedicated computer. The total number of tracks for each measured detector can be automatically saved according to the characterized identification code of detectors. Based on equation (1), the average radon concentrations were calculated with measured net tracks density, the exposure time and the appropriate calibration coefficients provided by National Institute of Metrology.

# 3. Results and Discussion

## 3.1. Optimum Etching Condition

To obtain the optimal etching conditions for these batches of detectors, 13 groups of experiments were conducted at the laboratory in Tsinghua University. Throughout the experiments, the given etching solution condition of the 6.25 mol/L sodium hydroxide (NaOH) aqueous solution remained unchanged and a combination of etching temperature and etching duration was selected. The standard etching conditions recommended by the manufacturer is etching for 1 hour at 98°C for radon. Then, considering the property of CR-39 track detector material, the selected etching temperatures ranged from 70°C up to 98°C. And to avoid over-etching at a high temperature, the etching duration was shortened carefully as the etching temperature increased. The selected etching conditions were as follows:

- Etching duration of 6 h,9 h,10 h,12 hand 18 h at a constant etching temperature of 70°C.
- Etching duration of 6 h,7 h and 12 h at a constant etching temperature of 75°C.
- Etching duration of 6 h and 8 h at a constant etching temperature of 80°C.
- Etching duration of 1 h,2 hand 3 h at a constant etching temperature of 98°C.

According to the track density, track shape, clarity, area density, uniformity and etching efficiency, all etching results were compared. The track density variation with etching temperature and etching duration is shown in Fig.3(a). As can be observed, the track density of the detectors treated for 6 h at 75 °C is the highest. Fig.3(b) shows two images of the etched α-particle tracks in CR-39 detector under this etching condition and the *TASLIMAGE* evaluation system. As can be seen from the figure, the track shapes were easy to read and the characteristic structure remained unchanged. In addition, for the etching efficiency, the combination of 75°C and 6 h was the best choice for detectors. As a result, the optimized etching parameters for this batch of detectors were selected as etching with 6.25 mol/L NaOH solution at 75°C for 6 h.

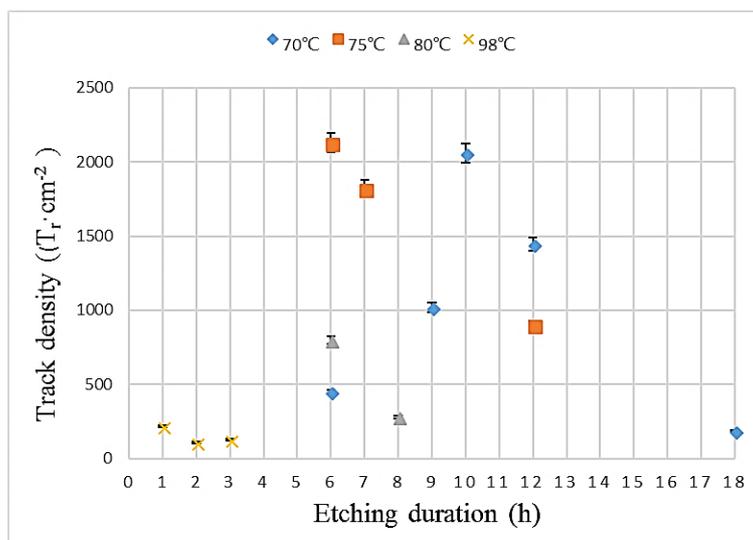

(a)

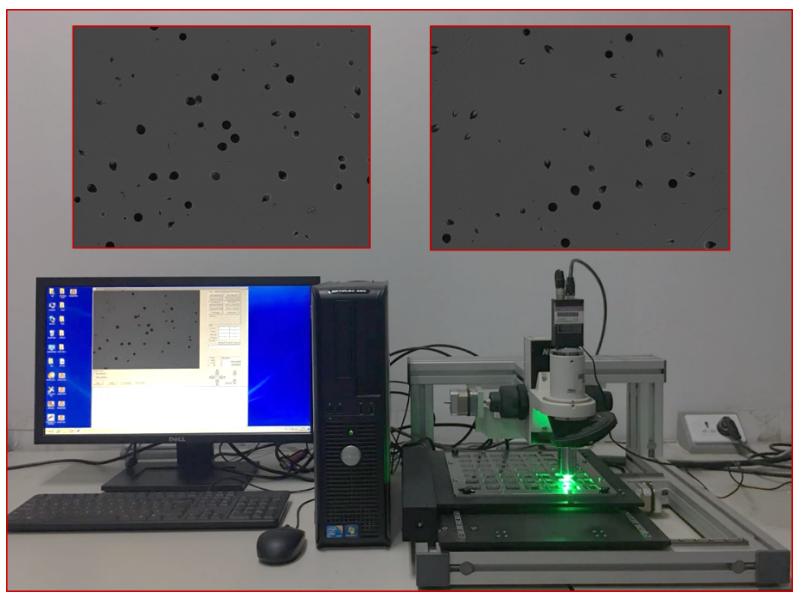

(b)

Fig.3.The etching results. (a)The track density variation with etching temperature and etching duration. (b)Two images of the etched α-particle tracks in CR-39 detector treated with 6.25 mol/L NaOH solution at 75°C for over 6 hours and the *TASLIMAGE* evaluation system.

## 3.2. Calibration Factor

Sixty detectors were placed in the radon chamber and different exposure tests were carried out at three cumulative radon concentrations of 544.585kBq·m$^{-3}$·h, 1440.907kBq·m$^{-3}$·h and 2845.295kBq·m$^{-3}$·h, respectively. The average calibration factor of these detectors was (1.0954±0.1068) Tr·cm$^{-2}$/kBq·m$^{-3}$·h and the relative standard deviation was 9.75%. Environmental parameters were maintained at 20°C and 40%RH throughout the experiment,

## 3.3. Radon Concentrations in CJPL-I

The measurement results at four sites of CJPL-I were shown in Table2.These measurements lasted for more than nine months and experienced nearly four seasons, avoiding radon concentration fluctuations caused by daily variation, seasonal changes and mechanical ventilation. They can be considered as the average annual radon concentration in CJPL-I.

In the PE room, the average annual radon concentration was (55±7) Bq·m$^{-3}$ and the radon concentration has a spatial variation ranging from 47 to 73 Bq·m$^{-3}$. The highest sampling point was located deep inside the room, and the lowest point was located in the doorway, which conformed to the influence of mechanical ventilation on radon concentrations distribution. Excluding the abnormal dates, the radon concentrations distribution was relatively uniform in space. A pair of parallel samples were placed on the shield and the radon concentrations were (57±4) Bq·m$^{-3}$ and (51±8) Bq·m$^{-3}$, respectively and the variable coefficient was 5.5%. Other parallel samples were placed on the distribution box and closed to the doorway with a variable coefficient of 4.1%.

In the main hall, the average annual radon concentration measured by the nuclear track detectors was (58±10) Bq·m$^{-3}$ and the spatial variations for radon concentrations were in the range from 40 to 77 Bq·m$^{-3}$ and the variable coefficients were 3.2% and 5.9%. The higher concentrations of the sampling points were 77 Bq·m$^{-3}$ and 74 Bq·m$^{-3}$, at the junction of the stairs and the first beam and far away from the ventilation, while the lower points of 40Bq·m$^{-3}$ and 45Bq·m$^{-3}$ were under the vent and close to the doorway. On the second floor of the hall, the average annual concentration was (53±9) Bq·m$^{-3}$ and the spatial variations for radon concentrations were in the range from 40 to 66 Bq·m$^{-3}$. There were several lower radon concentrations sampling points close to well-ventilated stairs. Other higher points were far away from the stairs entrance. The variable coefficients of parallel samples were 5.1% and 4.8%, respectively. As a result, compared with the measurement results in the period from 2010 to 2011, radon concentrations in the main experimental hall were reduced from (86±25) Bq·m$^{-3}$ (Mi et al., 2015) to (58±10) Bq·m$^{-3}$ due to the well-functioning ventilation system.

However, the average annual radon concentration in the storage tunnel was (345±15) Bq·m$^{-3}$. The spatial variations for radon concentrations were in the range from 314 to 365 Bq·m$^{-3}$ and the variable coefficients were 1.2% and 1.0%. Different from the other three sites in the experimental hall, the measurement results were significantly higher due to poor ventilation. There was no shield layer inside the storage tunnel and it was surrounded by bare rock without any ventilation system. Then, the measured results can be considered as the original radon concentrations in CJPL-I. Due to the relatively low radioactivity of marble contaminated by $^{232}$Th and $^{238}$U isotopes in the surrounding rock (Cheng et al., 2017), these levels of radon concentrations were lower than many underground structures (Li et al., 2006) and even inferior to the internal environmental pollution control specification for class II civil construction projects (400 Bq·m$^{-3}$). This can provide a more ideal radon concentration environment for low background and rare-event physical experiments.

Table.2. The minimum, maximum, average radon concentration and standard deviation at various sites in CJPL-I

| Sampling sites | The number of samples | $C_{Min}$(Bq·m$^{-3}$) | $C_{Max}$(Bq·m$^{-3}$) | $C_{Avg}$(Bq·m$^{-3}$) |
|---|---|---|---|---|
| The PE room | 12 | 47±8 | 73±10 | 55±7 |
| The main hall | 12 | 40±9 | 77±11 | 58±10 |
| The second floor of the hall | 12 | 40±8 | 66±10 | 53±9 |
| The storage tunnel | 12 | 314±16 | 365±12 | 345±15 |

The comparison between radon concentrations in the experimental hall of CJPL and other underground laboratories are reported in Fig.4. The average annual radon concentrations in CJPL were lower than those in Sudan((Thomas and D. M. Mei, 2012), Sudbury(Lawson, 2015), Sanford(J. B. R. Battat A et al., 2014; Thomas and D. M. Mei, 2012)and Canfranc(Bandac et al., 2017), while higher than those of Boubly(J. B. R. Battat A et al., 2014) and Gran Sasso(AF. et al., 2000; Bassingani et al., 1997) and similar to Y2L(MyeongJae Lee and Sun Kee Kim, 2011).It is shown that radon concentrations in CJPL-I were at an average level of those in underground laboratories in the world. The difference of radon concentrations in these underground laboratories depended on surrounding rocks, ventilation facilities, temperature, humidity, etc. The optimal ventilation was the main reason for low radon concentration in underground laboratories.

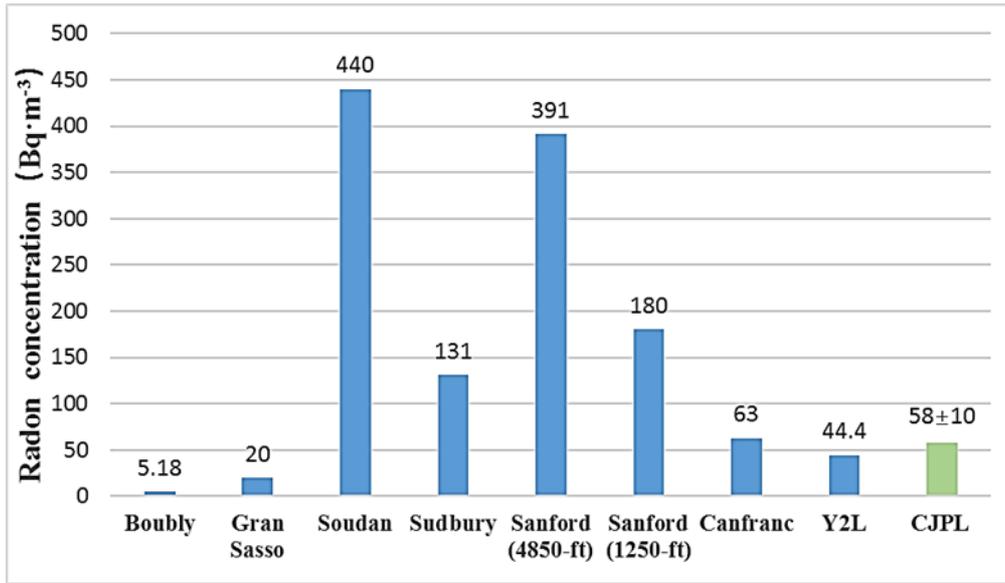

Fig.4. Comparisons of radon concentrations between CJPL-I and other underground laboratories in the world.

## 4. Conclusion

In this paper, radon concentrations in China JinPing underground Laboratory were measured using the solid-state nuclear track detectors within 290 days from May 30,2015 to March 16,2016. Three sites in the experimental hall were equipped with a well-functioning ventilation system with average annual radon concentrations of (55±7) Bq·m$^{-3}$, (58±10) Bq·m$^{-3}$ and (53±9) Bq·m$^{-3}$ respectively, all lower than the measured results during the commissioning period of ventilation equipment. In the storage tunnel without radon reduction systems, the radon concentrations rose to (345±15) Bq·m$^{-3}$, reflecting the original average annual radon concentration levels in the underground laboratory. In addition, the spatial variations for radon concentrations at four sites were also presented. At each site, radon concentrations at different sampling points varied with the ventilation conditions. The sampling points with lower radon concentrations were located in well-ventilated areas, and those with higher radon concentrations were in the deep interior of the room with poor ventilation conditions.

Compared with other international underground laboratories, the average radon concentration in CJPL-I was at an intermediate level. This radon radiation environment can meet the demand for low background and rare-event physical experiments. To further reduce radon radiation gas involved in low background experiments, more radon reduction methods will be adopted in the future, such as the planned low-radon air system.

## Acknowledgments

This work was supported by the National Key R&D Program of China (No. 2017YFA0402200, 2017YFA0402201), the National Natural Science Foundation of China (No.11675088, 11355001, 11175099), and Tsinghua University Initiative Scientific Research Program (No.20151080354). Thanks for the help of Mingkun Jing (Jinping Lab) and Juncheng Liang (National Institute of

Metrology).